# To Phrase or Not to Phrase – Impact of User versus System Term Dependence Upon Retrieval

Christina Lioma[1], Birger Larsen[2], and Peter Ingwersen[2]

[1] Department of Computer Science, University of Copenhagen, Denmark (c.lioma@di.ku.dk)
[2] Department of Communication, Aalborg University in Copenhagen, Denmark

**Abstract.** When submitting queries to information retrieval (IR) systems, users often have the option of specifying which, if any, of the query terms are heavily dependent on each other and should be treated as a fixed phrase, for instance by placing them between quotes. In addition to such cases where users specify term dependence, automatic ways also exist for IR systems to detect dependent terms in queries. Most IR systems use both user and algorithmic approaches. It is not however clear whether and to what extent user-defined term dependence agrees with algorithmic estimates of term dependence, nor which of the two may fetch higher performance gains. Simply put, is it better to trust users or the system to detect term dependence in queries? To answer this question, we experiment with 101 crowdsourced search engine users and 334 queries (52 train and 282 test TREC queries) and we record 10 assessments per query. We find that (i) user assessments of term dependence differ significantly from algorithmic assessments of term dependence (their overlap is approximately 30%); (ii) there is little agreement among users about term dependence in queries, and this disagreement increases as queries become longer; (iii) the potential retrieval gain that can be fetched by treating term dependence (both user- and system-defined) over a bag of words baseline is reserved to a small subset (approximately 8%) of the queries, and is much higher for low-depth than deep precision measures. Points (ii) and (iii) constitute novel insights into term dependence.

## 1 Introduction

When submitting queries to information retrieval (IR) systems, users may often specify which, if any, among the query terms are heavily dependent on each other and should be treated as a fixed phrase, for instance by placing them between quotes. The IR system then adapts the processing accordingly to retrieve text containing the same terms in the same order as what is inside the quotes. In addition to such cases where users specify term dependence, there also exist automatic ways for IR systems to detect dependent terms in queries [4, 21, 24] (overviewed in Section 2). Most IR systems support both such user and algorithmic approaches to detect term dependence in incoming queries. It is not however clear how much user and algorithmic assessments of term dependence agree, nor which of the two is likely to benefit retrieval performance the most.

To study this, we compare user assessments of term dependence to algorithmic assessments in 334 queries. We collect the user assessments by recruiting 101 search engine users through the CrowdFlower crowdsourcing platform and by examining their selection of term dependence. We produce the algorithmic assessments using four state-of-the-art term dependence ranking models [21,23]. Given a query, both user and algorithmic approaches decide if the query contains heavily dependent terms that should be treated as a fixed phrase instead of a bag of words. We compare retrieval performance between user and algorithmic methods of deciding term dependence, and also against a bag of words (no term dependence) baseline, using standard TREC datasets. Our findings agree with prior work [8,9] that users disagree not only with the algorithmic methods, but also among themselves. In addition, we report novel and interesting findings, showing for the first time, that this disagreement varies across different retrieval aspects such as query length or evaluation rank depth. Specifically we find that (i) there is little agreement among users about term dependence in queries, and this disagreement increases as queries become longer; (ii) the potential retrieval gain that can be fetched by treating term dependence over a bag of words baseline is reserved to a small subset of the queries, and is much higher for low-depth than deep precision measures.

We next overview related work (Section 2), and describe our crowdsourcing (Section 3) and retrieval (Section 4) experiments. We conclude by discussing our findings (Section 5).

## 2  Related Work

The user option of specifying term dependence in queries has existed since the mid 1970s as phrase operators, where a mixture of controlled vocabulary (descriptors), which contained phrases, and free-text searching was applied. Phrase (or proximity) operators have been particularly important in bibliographic IR systems, such as DIALOG or Web of Science. At the time, the users of bibliographic IR systems were mostly professional librarians, trained in using a wide range of operators including phrasing (spanning a range of term nearness options) from adjacent to a distance of $n$ terms in specified search fields, like title or abstract, or in the basic index. Early analyses of retrieval interaction from the late 1970s, e.g. [5, 27, 36], did not publish statistics on the use of phrase operators, but rather focused on the number and nature of query terms, eventually reaching the consensus that phrase and other proximity operators "were scarcely used". For instance, Fenichel [5] reported that novice users on average used 7.9 terms per search session, including descriptor terms, while moderately experienced users used on average 9.6 terms per search session, and experienced users used on average 14.4 terms per search session. Fenichel attributed the use of descriptor terms and phrases to the need for term alternatives and support during search. A decade later, Fidel [6] measured for the first time the use of phrase opera-tors by professional experienced DIALOG users, and found that (a) each command application, named a "move", applied 13.3 terms per search, and that (b) phrase operations constituted only 1.45% of all queries ([6], p. 518, Table 2). Other log analyses of the DIALOG system [11, 31, 33] also studied the use of phrase operators, among other things, but did not give statistics on their use per search and query *cycles*, nor on differences between novice and experienced searchers. No statistical evidence on the number of search terms per query *statement* or *cycle* was given, nor their nature (single terms, phrases, etc.).

Soon after, Web search was underway. Web search was done in a less structured environment than bibliographic IR: Up to approximately the year 2000, bibliographic IR mainly gave access to metadata in fields and an abstract per record; full records were introduced later. Another major difference between Web Search and bibliographic IR is that, although still fundamentally following Boolean logic, Web search did (and still does) not allow for set manipulation, did not have thesaurus support, and search sessions were overall shorter. As descriptors from a controlled vocabulary or a thesaurus (in bibliographic IR) leave little room for generating meaningful phrases and applying phrase operators, one would expect the use of phrase operators to increase in web search, compared to bibliographic search, practically leading to shorter search sessions than bibliographic retrieval. In addition, shorter search sessions, combined with a mostly layman rather than trained professional user population, is likely to have had an impact on the use of phrasing operators when searching. Indeed, that was the case. The first log-based study of web searching [13] studied 51K queries posed by 18,113 Excite users, where fixed phrases could be specified between quotes, and found that 6% of all users used phrase operators. Jansen et al. [13] suggest that the users had great difficulty in applying logical and language-based operations on the web from its start. Similar results to [13] were reported by Silverstein et al. [26], who studied 153M AltaVista queries, and by Spink et al. [34], who studied 531K EXCITE queries. Wang et al. [35] conducted the first longitudinal Web log study by analyzing 4 years' logged queries (541K) in a university website during 1997 -- 2001, and reported that some users were capable of querying using fixed phrases, but did not give statistics. Slightly later (2005), Jansen, Spink & Pedersen [14] studied 1.5M queries logged from AltaVista during 24 hours in 2002. They found that Boolean language was used for 6% of queries (p. 563), but no specific analysis of phrase operations was done. Among the 25 most frequent queries, none of them contained phrases. Similarly, Jansen, Booth & Spink did not analyse the phrase issue in their very large scale web log study carried out in 2005 [12]. They analysed 1,523,793 queries executed on the Dogpile meta search engine and found the average number of search terms per query reaching 2.79 *SD*=1, 54 terms (p. 1365, Table 4). Almost 71 % of the search sessions consisted of only one query.

The above log studies include few direct analyses of how users perceive term dependence in queries through phrase operators. It seems that overall users rarely query using fixed phrases: phrases have been used in queries at a rate of 1.45% in bibliographic retrieval [6] and 6% in web retrieval [13]. For the vast majority of queries, users tend to apply single terms as tokens for concepts [14, 35]. The use of the phrase operator seems to make more sense in free text search, where users must formulate the relevant phrase themselves. The number of terms in queries is difficult to establish, but we know the average number of query terms per bibliographic search session over the period 1980-97, as shown in Figure 1. If a search session on average consists of three iterations, corresponding to query submissions, and 15 search commands, including field codes and other logical operations [31], then each query on average consists of approximately 5 command operations. The overall small number of terms per query in bibliographic retrieval that we see in Figure 1 somewhat corresponds to the small term quantity later observed in free text retrieval on web engines. In web retrieval, the trend from the mid-90s is a slight growth in the use of multiple-term queries and thus an increase in the average number of terms applied per query, from 2.0 to 2.73 in 2005. However, no descriptors exist, and the searcher, most often inexperienced, must formulate his/her own query statements. One observes a large proportion of errors and scarce use of the phrase operator, 6% over all queries [13].

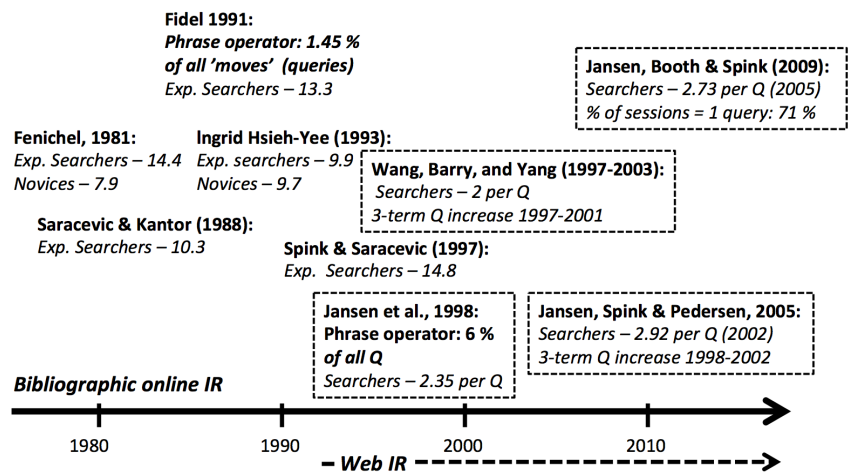

**Fig. 1.** Average number of query terms per search session in bibliographic online systems and terms per query statement in Web retrieval 1981-2005 based on different transaction-log analyses.

The reasons why phrases have been used so infrequently in information retrieval as a whole have not been studied. It is not clear to what extent users do not use phrasing because they think that it does not improve retrieval, or because they do not know of its existence, or because they cannot operate it properly, or because they tend to rarely apply meaningful term phrases when searching. Instead, users tend to apply single terms as tokens for concepts [14, 35]. Bibliographic searchers as well as web searchers appear to commit many errors, and failure may create uncertainty and lead to very simplistic query structures.

Even though in computational linguistics user assessments of term dependence have revealed interesting findings [25], for instance with respect to their lack of symmetry (cf. section 6 in [16], or e.g. larger dependence from *Pyrrhian* to *victory* than from *victory* to *Pyrrhian*) [24]), these advances have not been used in IR yet. In general, the last study recording how users specified term dependence was from 2005 [14].

On the contrary, algorithmic approaches to detect and process term dependence have been explored much more in IR, for instance in ad-hoc retrieval [20], patent retrieval [15], domain-specific retrieval on physics academic literature [17], or more formally using logic [19]. A recent comprehensive overview is given in [21]. It seems that the most popular methods for automatically detecting heavily dependent terms in queries rely on the co-occurrence frequency of the query terms in some query log or other large enough corpus (this has also resulted in thorough investigations of query term distributions [28]). The main premise is that the more often some query terms co-occur, the more dependent they are likely to be. This premise has been long applied [4, 26]. Recently, an alternative family of models was proposed [21] to automatically detect heavily dependent terms, which relies not on their frequency, but on their semantic distance when perturbed with synonyms. We use two of the best performing models of [21] to automatically detect heavily dependent terms in queries in Section 4.

Very relevant to our work is also the area of query segmentation or phrase identification, where several studies compare human versus automatics approaches to query segmentation and discuss their the impact on TREC data. One such large dataset for instance is published by Hagen et al. [9] and contains 50,000+ queries segmented by 10 annotators each. This dataset was subsequently also used by Hagen et al. again in [8]. In both papers, findings indicate low human agreement for some queries. Another dataset is published by Roy et al. [30] and contains 500 queries and 3 annotators. Further studies of human phrase detection have also been published in the NLP/Computational Linguistics community, see e.g. the work by Ramanath et al. [29], and the datasets of human segmentation by Bendersky, Croft & Smith [1], and Bergsma & Wang [2].

## 3 Crowdsourcing Term Dependence

To obtain human assessments of term dependence, we engaged 101 Web search engine users through the CrowdFlower[1] (CF) crowdsourcing platform. The CF experiment was entitled *To Phrase Or Not To Phrase -- Exact Phrases in Search Engine Queries* and included an initial *task description* phase (Section 3.1), a *training session* (Section 3.2), and the final *assessment session* (Section 3.3). We describe these next.

### 3.1 Initial Task Description

Users were introduced to the concept of quotes as exact phrase markers in queries, in order to receive results that contain that exact phrase and are potentially more accurate. They were then informed that they would be presented with queries and would have to select if and how to use quotes to specify exact phrases in those queries. They were asked not to use search engines to assess the results, but instead to decide based on their intuition and experience in web search. Figure 2 (a) shows the example shown to users, which illustrates all possible term dependence combinations for a query. The last option (*I do not understand the query*) was to be chosen when a query did not make sufficient sense for them to recommend whether to use term dependence or not. Only one option could be chosen per query.

---

[1] www.crowdflower.com

We showed only queries to users, not any associated context about the underlying information need or search task. On one hand, this may limit how well users understand the query and, by extension, how reliably they can assess if and when to specify fixed phrases in the query. On the other hand, this setup (of providing to users queries without any further information on the information need or search task) is a popular practice (Blanco et al. 2011 [3], Metrikov, Pavlu & Aslam 2015 [22], Yilmaz et al. 2012 [37]) that facilitates large-scale experimentation at relatively low cost (in IR experimental datasets, there exist significantly fewer queries with context information, than queries without context information). We chose the option of experimenting with a large number of queries, because the larger the query sample, the more generalizable and robust our findings on that sample. However, to address cases where users may not be able to understand the query due to lack of information on the underlying information need or task, we also specify the option "I do not understand the query". Users were instructed to use this option and simply skip queries they did not feel confident assessing.

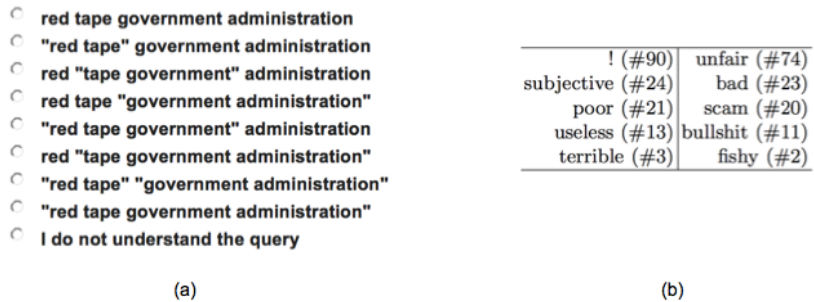

**Fig. 2.** (a) Example query with term dependence options given to CrowdFlower users. Quotes mark dependent terms. (b) 10 most frequent unigrams (with frequencies) extracted from user comments during training.

### 3.2    Session I: Training

The initial task description was followed by a compulsory training session on 52 test queries. Each of our 101 users was shown a query with all possible term dependence options, like in Figure 2 (a), and had to select one option only. Even though it would have made sense for users to be allowed to make more than one choice, the CF interface did not allow choosing multiple options. There was also a comment box for optionally typing feedback. After making their choice, users could see the answer we thought was correct, with an explanation. The queries used in this training session were not part of the queries used for retrieval later in Section 4, but a random selection from (a) the TREC 2012 Web adhoc track queries and (b) from queries that we made up to intentionally include heavily dependent terms in a majority of them. Table 1 displays the 52 train queries, with the most popular user choice of term dependence between quotes. Each of these 52 queries was assessed by 101 users. The scores in brackets in Table 1 show the average user agreement on the most popular user choice for each query, which we computed as the % of users (out of all 101 users) who agree on the most popular term dependence option for each query. For instance, the average agreement of 69% for "rain man" means that 70 out of 101 users ($\approx$ 69%) selected the option "rain man". The 52 train queries are sorted in Table 1 by decreasing user agreement.

**Table 1.** Train queries used on the CrowdFlower training session. Quotes mark the most popular user choice of term dependence for each query. Each query is assessed by 101 users. The percentages in brackets indicate how many out of the 101 users who assessed each query chose the most popular term dependence option (shown in this Table).

| | |
|---|---|
| *"rain man" (69%)* | *"george bush sr" bio (51%)* |
| *what is a "blue moon" (65%)* | *"bobbi kristina" funeral (51%)* |
| *what is a wiki (64%)* | *"down coats" canada (50%)* |
| *rules of golf (63%)* | *wineries niagara (50%)* |
| *grow peaches (63%)* | *presidential middle names (49%)* |
| *"doctor zhivago" (63%)* | *reviews of "les miserable" (49%)* |
| *generator for sale (63%)* | *how to get "windows 10" (47%)* |
| *what is madagascar known for (60%)* | *"frank lloyd wright" biography (47%)* |
| *"roosevelt island" (59%)* | *shipping cars from canada (45%)* |
| *world's biggest dog (59%)* | *"sacramento city college" (44%)* |
| *history of chile (58%)* | *average charitable donation (44%)* |
| *lump in throat (56%)* | *"grand bear lodge" coupon (43%)* |
| *"i will survive" lyrics (56%)* | *factory farming and poverty (43%)* |
| *afghanistan flag (56%)* | *"uss carl vinson" (42%)* |
| *"ford edge" problems (55%)* | *kids "earth day" activities (42%)* |
| *"nicolas cage" movies (54%)* | *"sweet potato" nutritional facts (41%)* |
| *"harry potter" birthday (54%)* | *hawaiian volcano observatories (39%)* |
| *it takes all sorts (54%)* | *usda food pyramid (38%)* |
| *"male menopause" (54%)* | *maryland department of natural resources (37%)* |
| *"lobster bisque" recipe (54%)* | *mens shoes size 13 (35%)* |
| *provinces of canada (54%)* | *"carpal tunnel syndrome" (35%)* |
| *inuit art (54%)* | *world war two germany leaders (35%)* |
| *"avant garde" etymology (53%)* | *antique dealer nautical (34%)* |
| *"dr samuel brown" atlanta (53%)* | *eggs shelf life (32%)* |

| *"raspberry pi" (53%)* | *"answering machine" messages from celebrities (24%)* |
| *"beef stroganoff" recipe (52%)* | *"athens airport" duty free (24%)* |

Table 2 summarises the statistics of the user assessments of the 52 train queries. User agreement in the last column of Table 2 refers to how many out of the 101 assessments received for each train query agree on the *single most popular* term dependence option for that query. We report the average of this number across all 52 queries (Table 2, row 5), or the average of this number across query subsets split according to query length (Table 2, rows 1-4).

We see in Table 2 that, overall, users disagreed on the most popular term dependence option for each query a bit more than they agreed (overall they agreed on average 49% of the times – see Table 2, row 5, last column). Comparing rows 1-5 in Table 2 we see that user disagreement increased as query length increased, probably because of the increased number of phrasing options (more query terms result in more term dependence combinations). We also see that users chose term dependence instead of bag of words in approximately 53.8% of all 52 train queries on average (Table 2, row 5). This rate is higher than what was reported in the literature in Section 2 because the 52 train queries were chosen to intentionally include term dependence in a majority of them. Note that allowing users to choose multiple options, which might be suitable for this task, was not allowed by the CF interface.

**Table 2.** Train queries statistics.

| row | query length | #queries | #bag of words | #phrasing | user agreement |
|---|---|---|---|---|---|
| 1. | 5 terms | 6 | 3 (50% of 6) | 3 (50% of 6) | 44.7% |
| 2. | 4 terms | 14 | 5 (35.7% of 14) | 9 (64.3% of 14) | 46.2% |
| 3. | 3 terms | 23 | 12 (52.2% of 23) | 11 (47.8% of 23) | 46.5% |
| 4. | 2 terms | 9 | 4 (44.4% of 9) | 5 (55.5% of 9) | 57.9% |
| 5. | sum or average | 52 | 24 (46.2% of 52) | 28 (53.8% of 52) | 48.8% |

Initially we planned to exclude users who failed the training session, as a way to combat crowdsourcing misconduct. Failing the training session consisted of disagreeing with the answer we thought was correct for 27 or more out of 52 train queries (i.e. more than half of the training queries). However, we soon observed that most users failed the training; in fact, they disagreed with our ground truth, just as much as they disagreed among them. This caused strong reactions from users, who described their frustration in the comments box. Figure 2 (b) shows the most frequent unigrams extracted from those user comments. Strongly negative adjectives and expletives prevail. We realised that this variation in user assessments was part of the subjective nature of this task, so we did not exclude users who failed the training. We did however filter users according to the user trust score that CrowdFlower provides, and selected only users with the highest user trust, as follows: CrowdFlower divides all users into three groups according to their trust score. CrowdFlower reports that this score is computed based on the user performance on previous tasks, but no further information on how this score is computed is given. The first group contains users of low user trust, the second of medium user trust, and the third of high user trust. We selected users from the third group only.

### 3.3 Session II: Assessment

After the training session, users proceeded to the assessment session. They were shown 20 queries per page, and had to select one term dependence option per query. We used 282 TREC queries and gathered in total 10 user assessments per query[2]. These 282 queries are all the TREC 6-8 queries (301-450, title only) of the AdHoc track and queries 1-200 of the Web AdHoc tracks of TREC 2009-2012, except those that contain only one term after stopword removal. Users had a minimum of 40 seconds to spend on each page, otherwise they were removed from the job. They were awarded 0.10 USD per page. We did not specify any maximum assessments per user, nor did we use restrictions on the crowdsourcing platforms that CF syndicates from, on geography, or on language. Even though users were asked to assess the queries without inspecting live Web search results, there is no guarantee that they did not do so. A pointer to this direction may be the time they spent on each assessment, which was overall quite low: all 2820 user assessments were completed in under 3 hours.

Table 3 summarises the user assessments of the 282 TREC queries in terms of query length, phrasing or bag of words choice, user agreement and trust. We explain user agreement and user trust next. In Table 3, user agreement refers to how many out of the 10 assessments received for each query agree on the *single most popular* term dependence option for that query. We report the average of this number across all 282 queries (Table 3, row 5), or the average of this number across query subsets split according to query length (Table 3, rows 1-4). User trust in Table 3 and Figure 3 is the trust score provided by CF as a number between 0 and 1 per user, and is based on user performance on training questions in previously completed jobs. There is no information on how this user trust score was computed. We report the average of this number across all 282 queries (Table 3, row 5), or the average of this number across query subsets split according to query length (Table 3, rows 1-4).

We see in Table 3 that, similarly to Table 2, users again tend to disagree a bit more than they tend to agree (average user agreement is 48.9% -- see row 5). Moreover, similarly to Table 2, user agreement also increased as query length decreased (see rows 1-4). However, unlike Table 2, users choose phrasing in approximately 34.9% of all queries. This is lower than the percentage we observed in the training queries because the queries in Table 3 were not chosen by us intentionally to include queries where phrasing was needed or according to whether they contained phrases or not. They are standard TREC queries.

---

[2] We collected 10 assessments per query. This does not imply that each user assessed 10 queries. Individual users assessed a different number of queries. Once we had 10 assessments per query, we removed that query from the pool of queries that were available for assessment in CrowdFlower.

**Table 3.** Query statistics (TREC queries).

| row | query length | #queries | #bag of words | #phrasing | user agreement | user trust |
|---|---|---|---|---|---|---|
| 1. | 5 terms | 10 | 8 (80% of 10) | 2 (20% of 10) | 40% | 0.55 |
| 2. | 4 terms | 22 | 14 (63.6% of 22) | 8 (36.4% of 22) | 45% | 0.53 |
| 3. | 3 terms | 120 | 79 (65.8% of 120) | 41 (34.2% of 120) | 48.6% | 0.51 |
| 4. | 2 terms | 89 | 56 (62.9% of 89) | 33 (37.1% of 89) | 62% | 0.52 |
| 5. | sum or average | 241 | 157 (65.1% of 241) | 84 (34.9% of 241) | 48.9% | 0.53 |

Figure 3 shows that higher trust users are more likely to use bag of words over term dependence, and vice versa (Spearman correlation coefficient ρ: 0.7). Sporadic use of term dependence actually gives better retrieval results, as we will see later in Section 4. So, it looks like more trusted users might be aware of this and might use term dependence more economically than less trusted users.

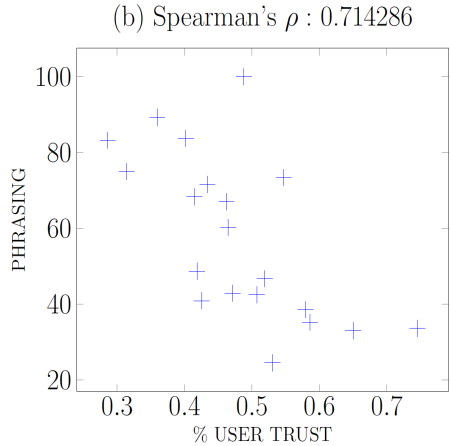

**Fig. 3.** CrowdFlower user trust (x axis) versus % of queries (out of all 282 queries) where users choose term dependence (y axis), binned.

Finally, we also computed the overall user agreement on *all* assessments (not only the most popular) to get a collective idea of the general agreement among our assessors. We used Krippendorff's alpha coefficient, which is a statistical measure of inter-annotator agreement that is applicable to any number of annotators, to incomplete (missing) data, and because it adjusts itself to small sample sizes [10]. Krippendorff's alpha coefficient *α =1* indicates perfect agreement, *α =0* indicates absence of agreement, and *α <0* indicates that disagreements are systematic and exceed what should be expected by chance. Krippendorff's alpha coefficient *α* is defined as follows:

$$a = 1 - \frac{Do}{De} \qquad (1)$$

where *Do* is the disagreement observed and *De* is the disagreement expected by chance.

$$Do = \frac{1}{n} \sum_{c \in R} \sum_{k \in R} \delta(c,k) \sum_{u \in U} m_u \frac{n_{cku}}{P(m_u, 2)} \qquad (2)$$

where $\delta$ is a metric function, $n$ is the total number of pairable elements, $R$ is the set of all possible annotations an annotator can give, $u$ is the annotations of all annotators for a given example, $U$ is the multiset of all $u$ for all examples, $m_u$ is the number of items in $u$, $n_{cku}$ is the number of $(c,k)$ pairs in $u$, and $P$ is the permutation function.

$$De = \frac{1}{P(n,2)} \sum_{c \in R} \sum_{k \in R} \delta(c,k) P_{ck} \qquad (3)$$

where $P_{ck}$ is the number of ways the pair $(c,k)$ can be made.

$$P_{ck} = \begin{cases} c \neq k & n_c n_k \\ c = k & n_c(n_c-1) \end{cases} \qquad (4)$$

Different metric functions $\delta$ can be used. Generally, for values $v$ and $w$,

$$\delta(v,w) \geq 0, \quad \delta(v,v) = 0, \quad \delta(v,w) = \delta(w,v) \qquad (5)$$

We found that the assessments of our users have in general a very low Krippendorff's alpha coefficient: *α <0.09*. This value of *α <0.09* means that disagreement among user assessments is too systematic to be by chance, hence that our findings are statistically generalizable.

## 4    Retrieval with User versus System Term Dependence

We compare the human assessments of term dependence collected in Section 3 (*user choice of term dependence*) to *automatic decisions of term dependence*, with respect to the retrieval performance they yield for the 282 TREC queries described in Section 3. Specifically we compare 1 run of user choice of term dependence to 6 runs of automatic decisions of term dependence. We explain these 7 runs below:

User choice of term dependence run:
(1) **User choice** of bag of words or term dependence per query. For each query, we use the most popular user choice (among the 10 CrowdFlower users). This choice can be either bag of words or any combination of term dependence, as the example in Figure 2(a) illustrates.

Automatic decisions of term dependence runs:
(2) **Bag of words** (no term dependence) for all queries.

(3) Automatic choice of bag of words or term dependence per query using the **Good Turing** (GT) model with median smoothing from [21]. (We explain this model in Section 4.1).

(4) Automatic choice of bag of words or term dependence per query using the **ATC** model from [21]. (We explain this model in Section 4.1).

(5) Treat as dependent only adjacent query terms, for all queries, using the well-known Markov Random Field model of sequentially dependent query terms (**MRF_S**) [23].

(6) Treat as dependent all query terms, for all queries, using the well-known Markov Random Field model of fully dependent query terms (**MRF_F**) [23].

(7) Choice of bag of words or any combination of term dependence among the query terms (as illustrated in Figure 2(a)), per query, according to what gives the best performance each time. This is an **upper bound** run, included to show the margin for improvement we can expect to achieve by selecting a bag of words or term dependence each time.

Next, we explain the GT and ATC models used respectively in runs (3) – (4).

For all seven runs, the ranking model is a unigram, query likelihood, Dirichlet-smoothed language model. We implement term dependence using the Indri query language for ordered windows *#1(...)*. For example, *#1(white house)* matches *white house* as an exact phrase. We use no stemming and remove stopwords from the queries only (as in [23]). We use Indri 5.8 for indexing and retrieval of at most 1000 documents per query. We evaluate retrieval effectiveness using four standard measures of low-depth and gradually deeper precision: MRR, P@10, NDCG@20, and MAP. We report these four evaluation measures for all 282 queries, not separately per query. We retrieve documents from Disks 4-5 (minus the Congressional Records for TREC7-8) for queries 301-450 and from ClueWeb09B for queries 1-200.

### 4.1   The GT and ATC term dependence models

GT and ATC detect which queries are more likely to be non-compositional. Non-compositional queries are queries whose meaning cannot be deduced from the meaning of their composing terms, such as *hot dog* or *red tape*, for instance. These queries must be treated as fixed phrases in IR [21]. GT works as follows:

Step 1.   It generates 'perturbed' queries, where a single query term at a time is replaced by a synonym

Step 2.   It produces a language model for each term in the original query and in each perturbed query (using distributional semantics of that term, extracted from some large corpus);

Step 3.   It combines the language models of the query terms to produce a language model for each query and for each perturbed query.

Step 4.   It computes the divergence between the language models of (a) the query and (b) its perturbed queries; the higher this divergence, the more non-compositional the query. Lioma et al. 2015 show that retrieval performance improves when non-compositional queries (detected in the above way) are submitted to the IR system inside quotes (i.e. are treated as fixed phrases of strong term dependence).

The GT model builds the language model of each query term (step 2 above) as follows:

$$P_{GT}(q,t) = \frac{(r+1)S(ff_{r+1})}{C_q S(ff_r)} \quad \text{for } r>0 \tag{6}$$

where $P_{GT}(q,t)$ is the probability of a term $t$ with frequency $r$ in query $q$, $ff$ is a vector with frequencies for term frequencies (also known as *double counts*), $Cq$ is the count of all terms in the context windows of $q$, and $S$ is a function fitted through the observed values of $ff$ to get the expected count of these values (see [7,21] for more). For zero count values, the probability is calculated as follows:

$$P_{GT}(q,t) = \frac{ff_1}{C_q} \quad \text{for } r=0 \tag{7}$$

where $ff_1$ is the frequency of frequency of *hapax legomena* (events occurring once). We extract the context windows from Disks4-5 for queries 301-450, and from ClueWeb09B for queries 1-200, exactly as described in Lioma et al. 2015. The above produces a language model for each term per query or perturbation. To produce one language model for the whole query or perturbation, we sort the language models of their terms and use the median of their values. We refer to this as GT median.

The ATC model follows the same high-level methodology as PG, with the difference that it produces vectors instead of language models for each query term in steps 2-3, and it computes the vector distance (instead of language model divergence) in step 4. Specifically, ATC builds a vector for each term (in step 2), where the elements of that vector correspond to that term's distributional semantics. The weight of each element in the vector is computed as the average of the weights of that term in all its context windows ($w_{it}$) as follows:

$$w_{it} = \frac{(0.5 + 0.5 \frac{f_{it}}{\max_f}) \log(\frac{N}{n(t)})}{\sqrt{\sum_{i=1}^{N} \left( \left(0.5 + 0.5 \frac{f_{it}}{\max_f}\right) \log(\frac{N}{n(t)}) \right)^2}} \tag{8}$$

where $w_{it}$ is the weight of term $t$ in context window $i$, $f_{it}$ is the frequency of $t$ in context window $i$, $max_f$ is the maximum frequency of any term in any context window, $N$ is the total number of context windows, and $n(t)$ is the number of context windows containing $t$. The vectors of all query terms are combined with their pointwise multiplication.

1414

### 4.2 Parameter Tuning

We tune parameter $\mu$ of the Dirichlet-smoothed ranking model in this range: {100, 500, 800, 1000, 2000, 3000, 4000, 5000, 8000, 10000}. We also tune the threshold $\theta$ of the Good Turing and ATC models, which controls how many queries to select as term dependent each time, identically to [21] in this range: {1 … 45} per TREC batch of 50 queries. To make sure that our results are not overfitted to the specific queries used in this experiment, we tune each parameter per evaluation measure using 3-fold cross validation, and we report the average of the three test folds.

### 4.3 Experimental Findings

Table 4 shows the results of our retrieval runs. When comparing user to system-selected term dependence, user selections are better for MRR on ClueWeb09, while system ones are better the rest of the times. User and system assessments agree 30.4% on average, meaning that it is the remaining 69.6% that impacts this behavior of MRR in ClueWeb09.

Comparing both user and system-selected term dependence to the upper bound, we see that users choose more term dependence (32% for Disks 4-5 queries and 38% for ClueWeb09 queries) than is actually required for optimal retrieval performance (between 1.6% - 6.6% for Disks 4-5 queries and between 9.3% - 17% for ClueWeb09 queries). The upper bound choice of term dependence is on average for all datasets and evaluation measures 8%. This value is much closer to the 6% reported in the literature for web search [10], than the user choice of term dependence which is on average (32% + 35%)/2 = 36.5%. This practically means that there is certainly margin for improving the automatic selection of term dependence through more strict selection. Interestingly, the users' intuitive, and possibly linguistic, interpretation of term dependence is the most damaging of all to retrieval performance.

The bag of words (BOW) run is mostly, but not always, the best method, across all datasets and evaluation measures. Note that bag of words was more often the choice of users with higher CF trust (cf. Figure 3). The reason why BOW tends to perform overall better than non-BOW approaches in our experiments may be connected to the distribution of query length in our dataset (shown in Table 5). Because we have removed 1-term queries from the TREC query sets we use, the majority of the queries (161 out of 282 queries in total) tend to have between 3 and 5 terms. The longer the queries, we reason, the more difficult it is to decide which part of the query, if any, has strong term dependence and should be treated as a fixed phrase. This difficulty was clearly shown in the human choice of term dependence reported in Table 3, where we see that user agreement on what part, if any, of the query should be placed between quotes decreases while query length increases, while user trust remains approximately the same. In most literature, experiments with TREC datasets are reported on the complete batches of queries, where the majority of queries contain 1-2 terms (i.e. they are relatively short). In those batches of queries, BOW is usually not the best performing method, because term dependence can be detected relatively more easily between 2 terms than between 3-5 terms.

**Table 4.** Retrieval effectiveness of manual (user-specified) versus automatic (algorithmically decided) term dependence. %UB is the % difference from the upper bound. %TD is the % of queries out of all 282 queries that use term dependence. USER CHOICE uses either bag of words or phrases, as chosen by manual user assessments (we adopt the most popular user choice on CrowdFlower). Out of the 6 automatic methods, MFR_S and MRF_F use term dependence for all queries, and GT, ATC use term dependence for a subset of the queries. UPPER BOUND uses bag of words or phrasing according to which fetches the best score. Bold in red boxes mark best scores (excluding the Upper Bound). N/A denotes Not Applicable.

| | | DISKS 4-5 | | | | | | CLUEWEB09B | | | | | |
|---|---|---|---|---|---|---|---|---|---|---|---|---|---|
| | | MAP | %UB | %TD | NDCG | %UB | %TD | MAP | UB% | TD% | NDCG | UB% | TD% |
| **USER CHOICE (MANUAL)** | | .1602 | -17.9% | 32% | .3732 | -17% | 32% | .1242 | -19.6% | 38% | .3928 | -12.7% | 38% |
| **SYSTEM CHOICE (AUTOMATIC)** | MRF_S | .1933 | -1.00% | 100% | .3983 | -14.6% | 100% | .1077 | -30.29% | 100% | .3463 | -23% | 100% |
| | MRF_F | .1833 | -6.0% | 100% | **.4341** | **-0.2%** | **100%** | .1151 | -25.5% | 100% | .3514 | -21.9% | 100% |
| | GT | .1949 | -0.2% | 73% | .4288 | -1.8% | 51% | .1168 | -24.4% | 56% | .3738 | -16.9% | 51% |
| | ATC | **.1950** | **-0.1%** | **77%** | .4290 | -1.8% | 55% | .1191 | -22.9% | 47% | .3811 | -15.3% | 55% |
| | BOW | .1933 | -1.00% | 0% | .4312 | -1.3% | 0% | **.1370** | **-11.3%** | **0%** | **.4052** | **-9.9%** | **0%** |
| | UPPER BOUND | .1952 | N/A | 6.6% | .4368 | N/A | 2.5% | .1545 | N/A | 12% | .4499 | N/A | 17% |
| | | P@10 | %UB | %TD | MRR | %UB | %TD | P@10 | UB% | TD% | MRR | UB% | TD% |
| **USER CHOICE (MANUAL)** | | .3778 | -14.5% | 32% | .6981 | -7.0% | 32% | .4921 | -13.8% | 38% | **.6028** | **-14%** | **38%** |
| **SYSTEM CHOICE (AUTOMATIC)** | MRF_S | .3687 | -16.6% | 100% | .6826 | -9.0% | 100% | .4120 | -27.8% | 100% | .5177 | -26.3% | 100% |
| | MRF_F | .4007 | -9.4% | 100% | .6957 | -7.3% | 100% | .4176 | -26.8% | 100% | .5050 | -28.1% | 100% |
| | GT | .4067 | -8.0% | 51% | .7321 | -2.5% | 51% | .4283 | -25% | 32% | .5568 | -20.7% | 51% |
| | ATC | .4053 | -8.3% | 56% | .7295 | -2.8% | 55% | .4308 | -24.5% | 53% | .5600 | -20.2% | 53% |
| | BOW | **.4377** | **-1.0%** | **0%** | **.7425** | **-1.1%** | **0%** | **.5011** | **-12.2%** | **0%** | .6000 | -14.5% | 0% |
| | UPPER BOUND | .4421 | N/A | 3.3% | .7508 | N/A | 1.6% | .5707 | N/A | 12.7% | .7021 | N/A | 9.3% |

However, even though BOW performs overall better than our other methods, we cannot conclude that phrasing may not be necessary, because of the upper bound reported in the last row: we see that BOW is performance-wise quite far from the upper bound, which uses sometimes phrasing and sometimes BOW (depending on which of them two fetches higher performance). Specifically BOW is between 1% and 14.5% worse than the upper bound, meaning that using BOW at all times, for all queries, is not the best choice.

Table 5. Query length of the 282 TREC queries.

| Query length | #Disks 4-5 queries | #ClueWeb09B queries |
|---|---|---|
| 5 terms | 0 | 2 |
| 4 terms | 7 | 13 |
| 3 terms | 76 | 63 |
| 2 terms | 56 | 65 |

We also see in Table 4 that the lower the depth of precision (i.e. measured by NDCG@20, P@10, MRR), the harder it is to approach the upper bound for ClueWeb09. We see this trend also in Figure 4, which shows that the highest gain is obtained for MAP and NDCG@20 when users agree approximately 70% and 90% respectively, whereas the highest P@10 and MRR gain is obtained when users agree approximately 20%. That is, improving low-depth precision is a much tougher task.

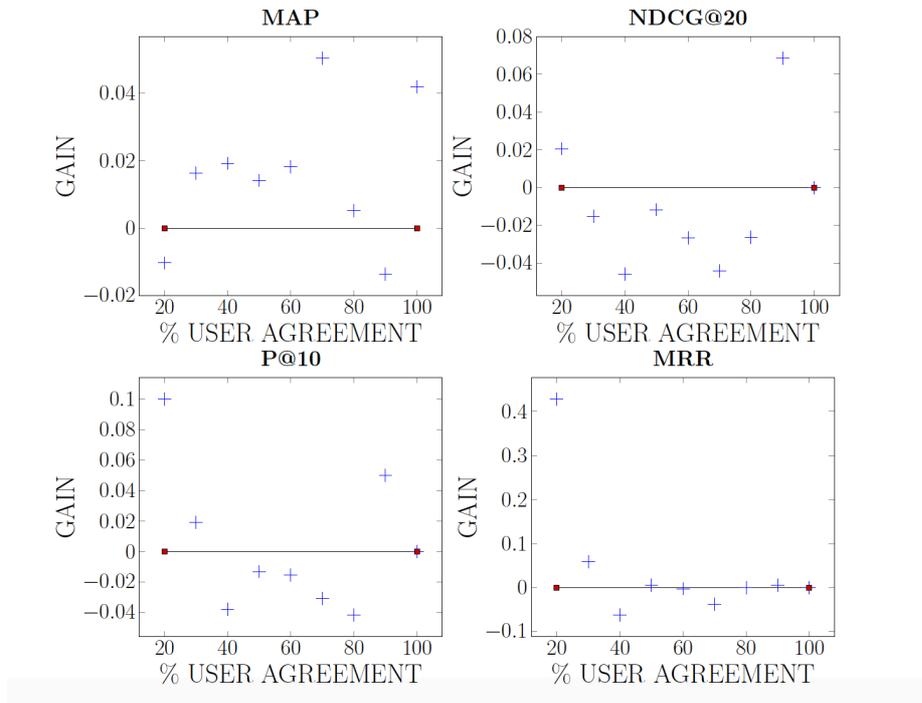

**Fig. 4.** Gain of user choice of phrasing or not over bag of words (y axis) versus user agreement (x axis). Binned. For positive y axis values, user choice > bag of words, and vice versa. The straight line marks no difference between user choice and bag of words.

To further understand the above results, we look at the top 5 queries where user choice outperformed the system choice, and vice versa (Tables 6 - 7). Several of the queries where user selection beats system selection tend to contain geographical names (*indiana, california, yelowstone, culpeper*). On the contrary, queries where system choice is best, tend to contain more high level descriptors that are more general and hence less discriminative in their meaning. Another reason why several queries where user choice underperforms compared to system choice could be the intuitive interpretation of a phrase by users, e.g. *british chunnel ..., schengen agreement, magnetic levitation ...*, without considering that treating these as fixed phrases may leave out synonymous or alternative phrases that are perhaps equally or even more frequent, such as *british channel tunnel, schengen treaty*. *Maglev* may have been misinterpreted by users as a proper noun, when in fact it is an abbreviation of *magnetic levitation*, and as such an alternative rather than part of the same phrase.

Table 6. Top 5 queries where user choice outperformed system choice. The best performing phrasing options per query are shown under column QUERY.

| TOP 5 QUERIES WHERE USER CHOICE OUTPERFORMS SYSTEM CHOICE |||||
|---|---|---|---|---|
| MAP || QUERY | NDCG || QUERY |
| USER | SYSTEM | | USER | SYSTEM | |
| .4043 | .0279 | "antibiotics inneffectiveness" | .3303 | .0000 | "indiana child support" |
| .4043 | .0279 | legionnaires disease | .5477 | .2547 | legionnaires disease |
| .2000 | .0134 | "culpeper national cemetery" | .4226 | .1331 | uplift at yellowstone national park |
| .1429 | .0281 | "indiana child support" | .3443 | .0589 | korean language |
| .2868 | .1722 | korean language | .3988 | .1299 | civil right movement |
| P@10 || QUERY | MRR || QUERY |
| USER | SYSTEM | | USER | SYSTEM | |
| .8000 | .1000 | korean language | 1.000 | .0000 | three gorges project |
| .6000 | .0000 | "california franchise" tax board | 1.000 | .0000 | anorexia nervosa bulimia |
| .6000 | .1000 | "er tv show" | 1.000 | .0033 | obama "family tree" |
| .9000 | .5000 | airport security | 1.000 | .0204 | law enforcement dogs |
| 1.000 | .6000 | "website" design hosting | 1.000 | .0244 | airport security |

Table 7. Top 5 queries where system choice outperformed user choice. The best performing phrasing options per query are shown under column QUERY.s

| TOP 5 QUERIES WHERE SYSTEM CHOICE OUTPERFORMS USER CHOICE |||||
|---|---|---|---|---|
| MAP || QUERY | NDCG || QUERY |
| USER | SYSTEM | | USER | SYSTEM | |
| .8369 | .1492 | schengen agreement | .7386 | .0000 | british chunnel impact |
| .6344 | .0561 | magnetic levitation maglev | .5887 | .0974 | income tax evasion |
| .4241 | .0921 | drug legalization benefits | .7846 | .0549 | hybrid fuel cars |
| .2649 | .0389 | viral hepatitis | 1.000 | .3639 | orphan drugs |
| .2413 | .0263 | hydrogen energy | 1.000 | .4694 | magnetic levitation maglev |
| P@10 || QUERY | MRR || QUERY |
| USER | SYSTEM | | USER | SYSTEM | |
| .7000 | .0000 | british chunnel impact | 1.000 | .0000 | kenmore gas water heater |
| 1.000 | .4000 | orphan drugs | 1.000 | .0000 | pacific northwest laboratory |
| .6000 | .1000 | hybrid fuel cars | 1.000 | .0000 | va dmv registration |
| 1.000 | .6000 | magnetic levitation maglev | 1.000 | .0000 | angular cheilitis |
| .6000 | .3000 | atypical squamous cells | 1.000 | .0000 | modern slavery |

## 5   Discussion and Conclusions

The main findings of our study of user decided vs. system decided term dependence are that i) there is little consensus among users about when to phrase query terms, ii) user-assessed term dependence differs significantly from algorithmically-assessed term dependence, and iii) the potential retrieval gain that can be fetched by any type of term dependence over a bag of words baseline is fairly low, but non-negligible with potential improvements possible in 8% of the queries. We also see that improving on low-depth precision is a much harder task, and that user decided term dependence for low-depth precision measures can outperform other approaches. As low-depth precision is important to users, this may explain why users use phrase operators in a small share of their searches as indicated in related work [14].

There are some limitations to our work. We use TREC queries, not users' own queries, and we evaluate retrieval with TREC relevance assessments, not by asking users. We do so for the sake of replicating and comparing to existing results. An explicit assumption of this, is that query phrasing can be perceived by users for a query that is not their own. For 99.9% of the assessed queries, users explicitly stated that they understood the queries they assessed. Even though understanding a query is not synonymous to cognitively formulating an information need and expressing it as a query, this study uses the former to approximate the latter, as is often common practice in such studies [18].

Furthermore, the low agreement among users about term dependence, combined with the CrowdFlower setup of only allowing one choice, meant that we could not use the training tasks as a quality filter as initially intended. The question is then if the quality of the crowdsourcing assessments is too low. We believe that most of the collected assessments are genuine, because (i) we chose users with the highest trust score provided by CrowdFlower, (ii) there is some agreement among users, and (iii) many complained about unfairness during training. Fraudulent users, we assume, are unlikely to spend extra time giving feedback (albeit negative) on the task. The fact that very few chose the *"I do not understand the query"* option indicates that there were no significant language fluency issues; if users were not fluent enough to understand a query, they would have skipped it.

Finally, even though we experiment with two standard TREC datasets containing 282 queries, and even though we make every effort to avoid overfitting by using x fold validation, our results may not be always generalizable to other data. We have chosen one dataset that is more representative of web search (ClueWeb09B) and one that is representative of more curated ad hoc search (Disks 4&5), but there are several other domains and contexts that are not represented in our experimental setup. We thus conclude that our findings are only reasonably valid for the domains represented by our TREC datasets.

In summary, in our experiments user defined term dependence improves retrieval performance in a minority of queries and mainly for low-depth precision. Some gains are possible for certain queries and the most promising direction to realise these improvements appears to be to focus on identifying these automatically, either statistically or by further developing linguistically informed methods such as those in [21]. In the future, we plan to investigate the effects of strength or degree of term dependence. We did not do so in this study, to keep our scenario similar to the real-life search scenario of using quotes to search for phrases. However, as the automatic approaches (GT and ATC from [21]) output a degree of term dependence, and as we have collected 10 assessments of user decided term dependence per query, in the future we plan to investigate the effect of degrees of term dependence.

**Acknowledgements**

This work has been partially funded by the following grants: REACT (Responsible Impact), supported by Det Obelske Familiefond Danmark; QUARTZ (Quantum Information Access and Retrieval Theory), supported by Horizon 2020 Marie Skłodowska-Curie Innovative Training Networks.